\documentclass[12pt]{article}
\usepackage{fancyhdr}
\usepackage{latexsym}
\usepackage{amssymb}
\usepackage{graphicx}
\usepackage{epsf}
\usepackage{amscd}
\usepackage{amsmath}
\usepackage{color}

\usepackage{amsthm}
\usepackage{mathrsfs}

\makeatletter
\@addtoreset{equation}{section}

\renewcommand{\thefootnote}{\fnsymbol{footnote}}
\makeatother

\def\ee{\end{eqnarray}}

\def\=:{=\hspace{-.7em}\raisebox{1.1ex}{.}\hspace{.1em}\raisebox{-0.2ex}{.} }
\newcommand {\beq}{\begin{eqnarray}}
\newcommand {\eeq}{\end{eqnarray}}

\begin{document}
\thispagestyle{empty}

\begin{flushright}
arXiv:{\tt yymm.nnnn} [hep-th]\\ %, \;\;
November, 2008 \\
\end{flushright}
\vspace{1mm}

\begin{center}
{\Large \bf
Colliding Waves on a Brane,\\
the Big Bounce and Reconnection}
\\[10mm]
\vspace{2mm}

{\normalsize\bfseries
Tae-Hun Lee$^{a}$ and Muneto Nitta$^b$
}
\footnotetext{
e-mail~addresses: \tt
lee109@physics.purdue.edu, nitta(at)phys-h.keio.ac.jp
}

\vskip 1.5em
$^a$ {\it Purdue University, Department of Physics, West Lafayette,
IN 47906, USA
}
\\
$^b$ {\it Department of Physics, Keio University, Hiyoshi, Yokohama,
Kanagawa 223-8521, Japan
}
\vspace{2cm}
%
%\vfill \eject
%\newpage
%

{\bf Abstract}\\[5mm]
{\parbox{13cm}{\hspace{5mm}
%%%%%%%%%%%%%%%%%%%%%%%%%%%%%%%%%%%%%%%%%%%%%%%%%%%%%%%%%%%%
We present a time-dependent solution of the Nambu-Goto action which represents two colliding waves moving at the speed of light.
This solution can be decomposed into two distinct regions
with different geometries corresponding to shrinking or
expanding brane universe and two colliding branes.
The former describes the Big Bounce without singularity while
the latter describes that two branes collide
and reconnect to each other.
The colliding brane region has a signature change.
Classical dynamics of a massive particle on the brane is studied through the geodesics.
%%%%%%%%%%%%%%%%%%%%%%%%%%%%%%%%%%%%%%%%%%%%%%%%%%

}}
\end{center}
\vfill
\newpage
\setcounter{page}{1}
\setcounter{footnote}{0}
\renewcommand{\thefootnote}{\arabic{footnote}}

%%%%%%%%%%%%%%%%%%%%%%%%%%%%%%%%%%%%%%%%%%%%%%%%%%%%%
\section{Introduction}\label{sec:Introduction}

A particular exact solution to a classical field equation is often regarded as to an extended particle under certain special conditions,
a so-called soliton (for a review,  \cite{Rajaraman}),
which is important in study of nonperturbative properties
of field theories.
The well-known example is a kink solution to
the sine-Gordon equation \cite{Perring:1962vs}.
Such a solution can contain not just one particle picture but multi-particle interaction process
\cite{Perring:1962vs, Belinsky}.
Recent examples of exact solutions of multiple kinks are
obtained in $U(N)$ gauge theories at strong coupling \cite{Isozumi:2004va}.
A time dependent solution which cannot be obtained by boosting a static solution can be nontrivial.
It is the case if a solution represents a massless particle or more than one particle interaction process.
In this paper we focus on solutions of the Nambu-Goto action
\cite{Nambu:1974zg} which describes fluctuations of a brane.
Some of exact static solutions of the Nambu-Goto action
or the Dirac-Born-Infeld action \cite{Dirac:1962iy,Born:1934gh} 
for a brane are well known. 
For example, the electric BIon solution(catenoid) is a well-known
static solution \cite{Born:1934gh,Callan:1997kz}. 
See \cite{Townsend:1999hi} and references therein
for more examples of static solutions. 
As an example of time-dependent solutions,
the Scherk's surface for colliding branes is known, 
in which two branes reconnect each other 
in collision \cite{Gibbons:2006rr}.

%%%%%%%%%%%%%%
On the other hand,
when solitons have linear structures,
propagating waves on them are
time-dependent solutions of the effective action,
which is typically the Nambu-Goto action.
For example,
waves propagating on vortex-strings and on domain walls
were previously studied \cite{Garfinkle:1990jq}.
Waves along strings in gravity \cite{Economou:1991bc} and
supergravity \cite{Garfinkle:1992zj}, black strings and D-strings \cite{wavy-strings} were also studied.
These are based on the fact that
the Nambu-Goto action admits wave solutions
with arbitrary shape, propagating at the speed of light;
For instance the Nambu-Goto action for a $p$-brane of codimension one
is given in the static gauge by
\begin{equation}
 S = \int d^{p+1}x \mathscr{L}
 = -\sigma\int d^{p+1}x \sqrt{1-\partial \phi(x) \cdot \partial \phi(x)}
 \label{eq:Nambu-Goto action}
\end{equation}
from which the equation of motion for the fluctuation $\phi(x)$
reads
\begin{equation}
 \partial^2\phi + \frac{1}{2[1-(\partial\phi)^2]}
 \partial \phi\cdot\partial\left[(\partial\phi)^2 \right] =0.
\label{eq:EOM}
\end{equation}
This admits wave solutions in an arbitrary shape
propagating at the speed of light into one space direction of the $p$-brane world-volume, given by
\begin{equation}
 \phi(x)=f(\vec{k}\cdot\vec{x}\pm\omega t+c),
 \label{eq:single wave}
\end{equation}
where $\vec{k}^2 -\omega^2=0$, $c$ is an arbitrary constant and $f$ is an arbitrary function.
In the case of Bogomol'nyi-Prasad-Sommerfield(BPS) solitons
in supersymmetric theories,
waves with an arbitrary shape, propagating at the speed of light
along vortex-strings, domain walls and 1/4 BPS composite states
(see, e.g.,\cite{Eto:2006pg})
have been shown to be still BPS \cite{Eto:2005sw}
(see also \cite{BlancoPillado:2006qu}).
To the best of our knowledge,
waves propagating to
only one direction of solitons have been explicitly known so far
as exact solutions.
If initially two waves are simultaneously prepared
in well separated regions,
for a while each of them will preserve the forms (\ref{eq:single wave}) without interference,
so that the configuration is approximately a superposition of them
if they are not much overlapped.
However, once those waves get close to each other
such configuration is no longer valid;
we should take into consideration the full non-linear equation (\ref{eq:EOM}),
which is a highly non-trivial problem, though some approximate solutions of two colliding waves
can be found in the literature \cite{Siemens:2001dx}.

In this paper we present an exact
solution of two colliding waves moving
at the speed of light in the Nambu-Goto action.
In our solution
two waves collide and scatter each other.
The solution in our interest is a linear combination
of the left and the right moving waves.
This is also a solution to non-interacting wave equation of
the Klein-Gordon type.
It turns out that the only nontrivial solution
is a special linear combination of two logarithmic waves
with singular peaks.
Our solution should be immediately applied to colliding
waves on solitons which have extended directions,
such as domain walls and strings.
It will be also relevant to dynamics of cosmic strings.

Our solution can be decomposed into
regions having two different geometries
which describe different physics.
The central region between the two peaks
has the Robertson-Walker type induced metric.
It describes shrinking and expanding universe
connected by ``Big Bounce" where no
``Big Bang" singularities exist.
The region outside the two peaks represents collision
dynamics of two branes moving at the speed of light.
In this case, two branes reconnect with each other in collision.
This solution is new and
different from the Scherk's surface for colliding branes
which also describes reconnection
\cite{Gibbons:2006rr}.
The geometry on the outside branes has
an Euclidean region where
the positivity of the inside of the square root of the action is violated.
It is quite interesting that a single continuous solution covers over
both Minkowskian and Euclidean regions.
Such a situation was already discussed in the literature
\cite{Gibbons:2004dz}. We also calculate the energy of the brane fluctuations and find
that it diverges due to the singular peaks.
%in Appx. \ref{appx:Energy of brane}.

A classical particle dynamics on the brane is
discussed in detail in two separate regions according to their geometries.
While a massive particle on the inside brane is on the scale changing universe, one on the outside branes is bounded by the potential.  In the end it is verified that a massive particle on the brane cannot move faster than the speed of light
in all the regions in the frame of bulk.

This paper is organized as follows.
In Sec.~\ref{sec:Wave solutions} we start to discuss wave solutions of the Nambu-Goto action and then move onto a particular solution in our interest, which is two waves moving at the speed of light in two different directions on the Nambu-Goto brane. There we discuss its properties in detail, before going into discussion of a massive particle dynamics on the brane.
The geometries of the inside region (the Big Bounce brane universe) and the outside region (colliding branes) are investigated through geodesics of a massive particle in Sec.~\ref{sec:Expanding brane universe} and Sec.~\ref{sec:Colliding branes}, respectively. Each of these sections ends with verification that a speed of a massive particle in the bulk frame cannot exceed the speed of light. Finally, Sec.~\ref{sec:Conclusion and Discussion} is devoted to conclusion and discussion. Divergence of the energy of the brane is shown in Appx.~\ref{appx:Energy of brane}.
%\vspace{1cm}

%\newpage
%%%%%%%%%%%%%%%%%%%%%%%%%%%%%%%%%%%%%%%%%%%%%%%%%%%%%%%%%%%%%%
\section{Wave solutions of the Nambu-Goto action}\label{sec:Wave solutions}
In this section
we look for further solutions to the equation of motion (\ref{eq:EOM})
of the Nambu-Goto action.
In the linear approximation up to the first order of $\phi$, only the first term in Eq.~(\ref{eq:EOM}) remains,
reducing to the Klein-Gordon equation, $\partial^2 \phi = 0$.
In this limit it admits linear waves
\begin{equation}
 \phi(x) = \sum_i f(\vec{k_i}\cdot\vec{x} \pm \omega_i t + c_i)
 \label{eq:linear sum}
\end{equation}
with $\vec{k_i}^2 -\omega_i^2=0$ and $c_i$ are arbitrary constants.
This approximation is valid when the brane
fluctuation is small enough.
However, when a description of large fluctuations is needed, to solve the full equation is highly nontrivial
because of nonlinearity of the second term.
Some of solutions above,
also satisfy the whole nonlinear equation (\ref{eq:EOM}).
The simplest example has been already shown in Eq.~(\ref{eq:single wave})
which consists of one wave.
Note that nonlinearity in general does not allow
a linear combination (\ref{eq:linear sum})
of wave equation solutions in different directions.
However, a certain particular type of linear combinations of waves is
found here.
For solutions to the linear wave equation,
$\partial^2\phi=0$, Eq.~(\ref{eq:EOM}) is simply reduced to
\begin{equation}
 \partial\phi\cdot\partial\left[(\partial\phi)^2\right]=0.
 \label{eq:log-EOM}
\end{equation}
On generalization of the solution (\ref{eq:single wave})
of waves propagating into one direction,
we consider an ansatz for two waves propagating at the speed of light
along two directions of
Lorentzian momentum vectors $k$ and $p$,
\begin{equation}
 \phi(x) = f(k\cdot x)+g(p\cdot x) .
\end{equation}
Substituting this to (\ref{eq:log-EOM}) yields
\begin{equation}
\begin{array}{ccl}
0&=&(k_mf^\prime +p_m g^\prime)\partial^m
[k^{2}f^{\prime 2}+p^{2}{g^\prime}^2
 + 2(k\cdot p)f^\prime g^\prime]\\
&=&2(k\cdot p)(k_mf^\prime +p_m g^\prime)
 (k^{m}f^{\prime\prime}g^\prime +p^{m}f^\prime g^{\prime\prime})\\
&=&2(k\cdot p) [k^{2}f^{\prime\prime}f^\prime g^\prime
 + p^{2}f^\prime g^\prime g^{\prime\prime}
 + (k\cdot p)(f^{\prime2}g^{\prime\prime}
 + {g^\prime}^2f^{\prime\prime})]\\
&=& 2(k\cdot p)^2\left[f^{\prime 2}(k\cdot x) g^{\prime\prime}(p\cdot x)
  +{g^\prime}^2(p\cdot x) f^{\prime\prime}(k\cdot x)\right].
\end{array}
\end{equation}
Here, $m$ runs over the world-vlume coordinates
of the brane, $m=0,1,\cdots,p$,
and a prime denotes differentiation of functions
with respect to their arguments.
Then, for $p \cdot k \neq 0$, we obtain
\begin{equation}
  \frac{f^{\prime\prime}(k\cdot x)}{f^{\prime2}(k\cdot x)}
=-\frac{g^{\prime\prime}(p\cdot x)}{g^{\prime2}(p\cdot x)}=-\frac{1}{s},
\end{equation}
where $s$ is an arbitrary real constant. Therefore, we obtain
\begin{equation}
 \phi(x)=s[\ln(k\cdot x+c_1)-\ln(p\cdot x+c_2)]+c_3
\end{equation}
where $c_1$, $c_2$ and $c_3$ are arbitrary constants.

\begin{figure}[h]
  \begin{center}
\begin{tabular}{cc}
    \includegraphics[scale=0.7]{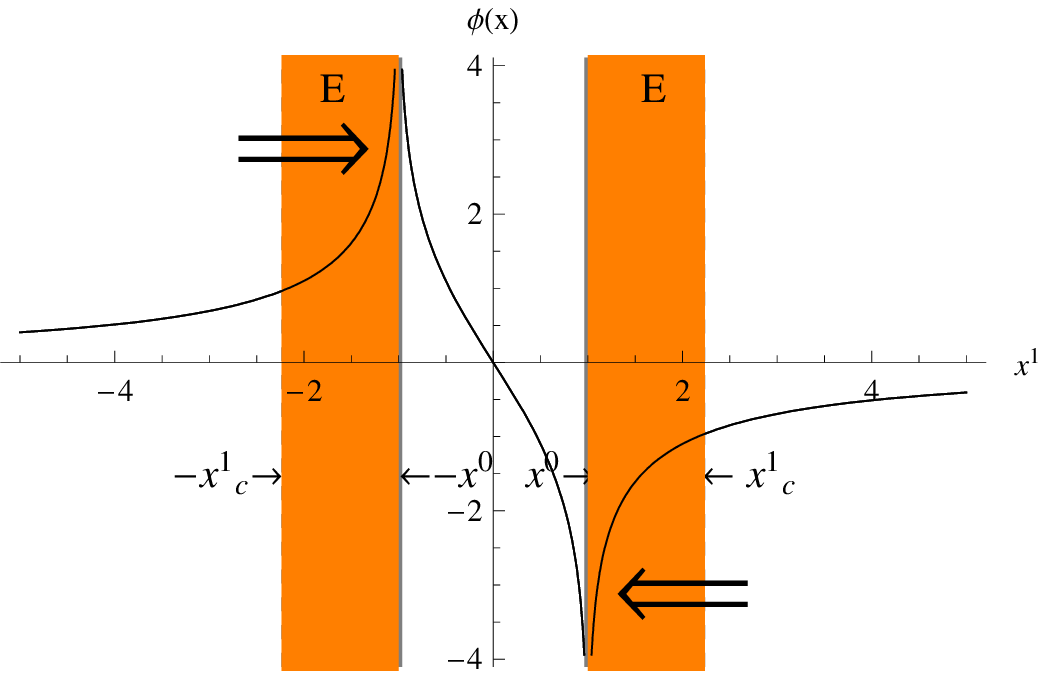} &
    \includegraphics[scale=0.7]{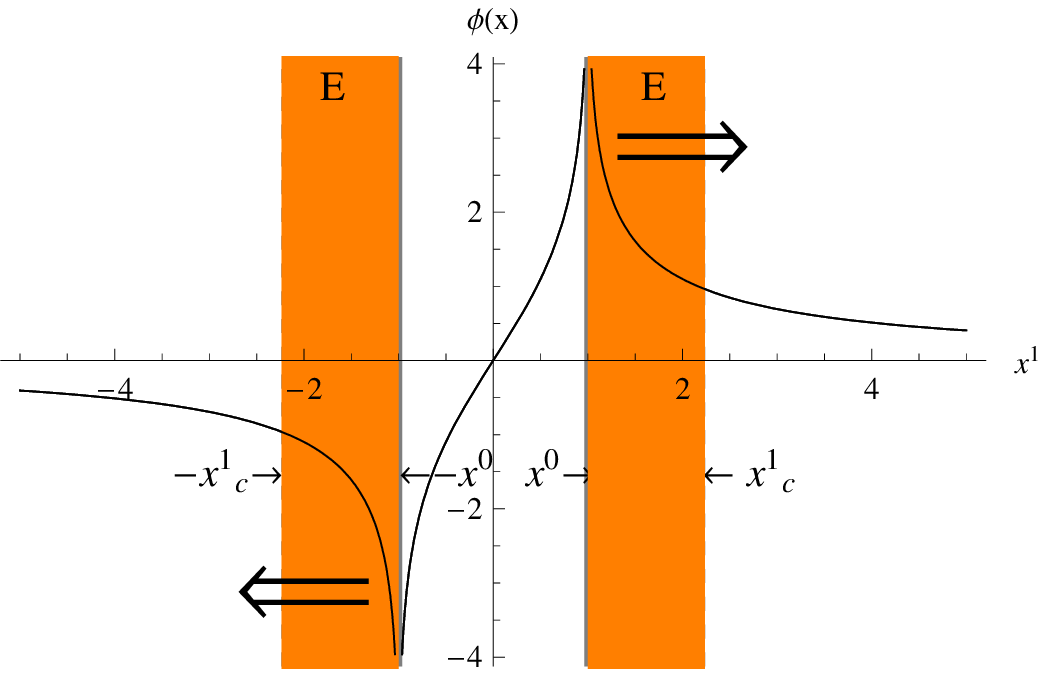}\\
(a) & (b)
\end{tabular}
\end{center}
\caption{
The solutions $\phi(x)$ with $s=1$ are plotted for
(a) at $x^0=-1$ and (b) at $x^0=1$.
The arrows represent the directions of moving two peaks.
The two shade rectangles marked by ``E"
denote the Euclidean regions, as explained below.
\label{fig:log-sol}
}
\end{figure}

We now consider the simplest case of
colliding waves, $\vec{p} = - \vec{k}$.
Without loss of generality,
by Lorentz transformation and translation,
the solution can be written as
\begin{equation}
 \phi(x)=s(\ln|x^0+x^1|-\ln|x^0-x^1|).
\end{equation}
See Fig.~\ref{fig:log-sol} for a profile of this solution.
%%%%%%%%%%%%%%%%%%%%%%%%%%%%%%%%%%%
%%%%%%%%%%%%%%%%%%%%%%%%%%%%%%%%%%
There exist two singular peaks at $x^1 = \pm x^0$
which move at the speed of light.
They collide and scatter each other at $t=0$.
Equivalently, it can be written for two separate regions as
\begin{equation}
\phi(x)=
\left\{
\begin{array}{l}
\displaystyle
 s\ln\frac{x^0+x^1}{x^0-x^1}; \mbox{ for } (x^0)^2>(x^1)^2:\mbox{I-Minkowskian},\\
\displaystyle
 s\ln\frac{x^0+x^1}{x^1-x^0}; \mbox{ for } (x^0)^2<(x^1)^2:\mbox{II-Minkowskian \& Euclidean}.
\end{array}\right.  \label{eq:separate solutions}
\end{equation}
It will be seen that each of the solutions has a different geometry.
We call the first region between the two peaks
as ``the Big Bounce brane universe''
and the second region outside the two peaks
as ``the colliding branes''.
We discuss these geometries
in the following sections.

Next, it is necessary to check
whether all the region is valid making the original action real.
Inside the square root of $\sqrt{1-(\partial\phi)^2}$
of the Nambu-Goto action (\ref{eq:Nambu-Goto action}),
\begin{equation}
\begin{array}{ccl}
1-(\partial\phi)^2&=&1-(\partial_0\phi)^2+(\partial_1\phi)^2\\
&=&1-s^2\left(\frac{1}{x^0+x^1}-\frac{1}{x^0-x^1}\right)^2+s^2\left(\frac{1}{x^0+x^1}+\frac{1}{x^0-x^1}\right)^2\\
&=&1+\frac{4s^2}{(x^0)^2-(x^1)^2}\geq0,
\end{array}
\end{equation}
is required in order to take a real value.
This condition can be rephrased as
\begin{equation}
\left\{
\begin{array}{l}
(x^0)^2>(x^1)^2\\
(x^1)^2\geq4s^2+(x^0)^2.
\end{array}\right.
\end{equation}
In other words, the region
\begin{equation}
 (x^0)^2 < (x^1)^2 < (x^0)^2 + 4s^2\equiv(x^1_c)^2 \label{eq:Euclidean}
\end{equation}
is somewhat pathological because the action becomes
purely imaginary.
We call this region as the ``Euclidean region".
The two Euclidean regions are shaded  and
are denoted by ``E" in Fig.~\ref{fig:log-sol}.

The first region I in Eq.~(\ref{eq:separate solutions})
(the Big Bounce brane universe) is purely Minkowskian.
But the second region II in Eq.~(\ref{eq:separate solutions})
(the colliding branes) contains
the Euclidean region (\ref{eq:Euclidean}) as well as the Minkowskian region.
If we could remove the Euclidean region from the solution,
no singularity would be included in the rest.
However, this problem is subtle because every space is connected by energy density flow.
The similar situation has already occurred in the literature
\cite{Gibbons:2004dz}.
In this circumstance the energy of the branes is calculated in Appx.~\ref{appx:Energy of brane}.
We find that
the energy of the Minkowskian part of the region II is finite
but the total energy is infinite due to
the region I and the Euclidean region in the region II.

In the next sections the dynamics and geometry of these two solutions will be discussed on the same outline. Before going into the discussion, it is necessary to mention that it is sufficient to consider only $x^0>0$ since the solution has the symmetry
\begin{equation}
x^0\rightarrow-x^0 \Leftrightarrow s\rightarrow-s.
\end{equation}

%%%%%%%%%%%%%%%%%%%%%%%%%%%%%%%%%%%%%%%%%%%%%%%%%%%%%%
\section{The Big Bounce brane universe}\label{sec:Expanding brane universe}
\begin{figure}[h]
  \begin{center}
\begin{tabular}{cc}
     \includegraphics[scale=0.7]{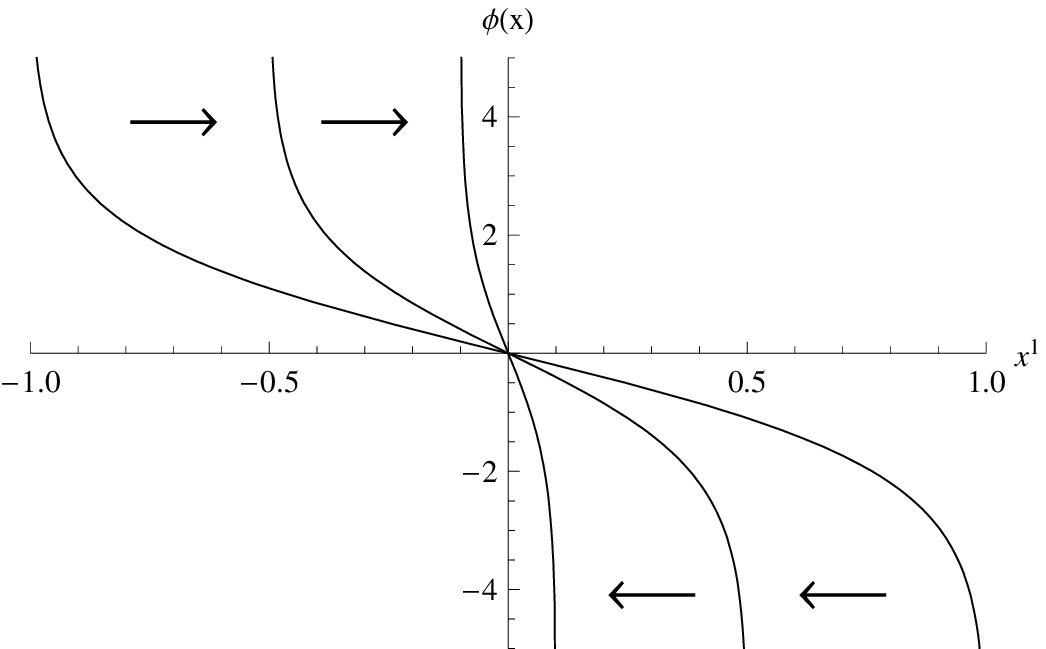}&
    \includegraphics[scale=0.7]{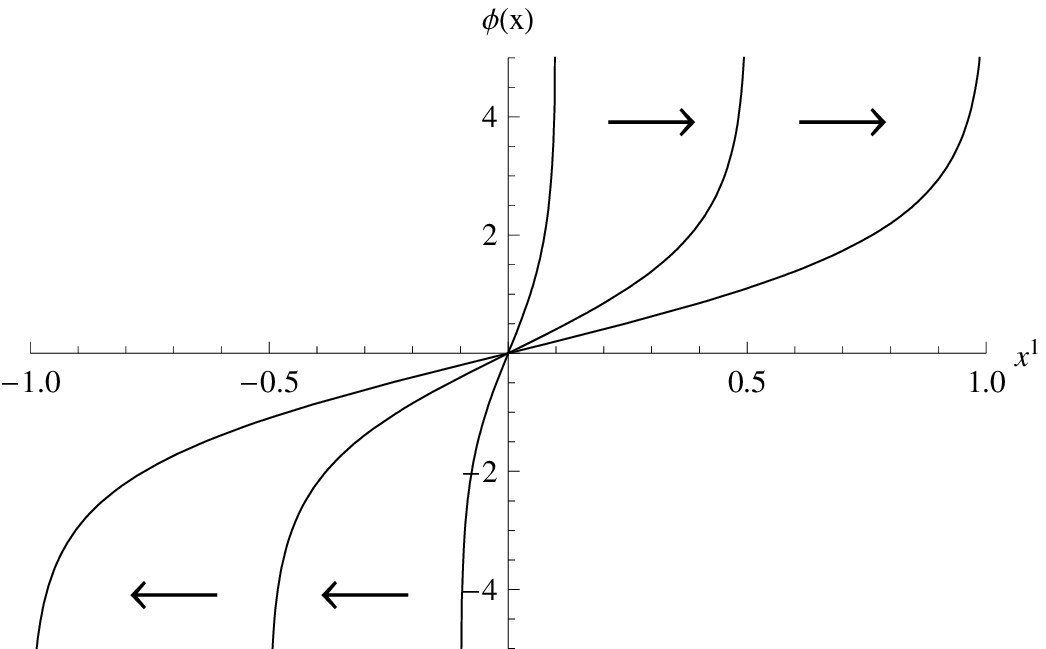} \\
 (a) & (b)
\end{tabular}
    \end{center}
\caption{The dynamics of the Big Bounce brane universe. (a) The branes at $x^0=-1, -0.5, -0.1$ for $x^0<0$ and $s=1$ in the bulk frame. The universe shrinks as time goes on while the brane itself becomes vertical.
(b) The branes at $x^0=0.1, 0.5, 1$ for $x^0>0$ and $s=1$ in the bulk frame.
The brane is stretched as time goes on.
The arrows represent the directions
which the branes are moving to.
\label{fig:brane_universe}}
\end{figure}

In this section we study the central region describing
the Big Bounce brane universe, see Fig.~\ref{fig:brane_universe}.

%%%%%%%%%%%%%%%%%%%%%%%%%%%%
%%%%%%%%%%%%%%%%%%%%%%%%%%%
The line element with induced metric on the $p$-brane of
codimension one in the static gauge is written as
\begin{equation}
 ds^2=\eta_{mn}dx^mdx^n-d\phi^2.
\end{equation}
Omitting the trivially extended directions of the line element
$\eta_{ij}dx^idx^j$, where $i,j=2,3,\cdots,p$, it can be written as
\begin{equation}
ds^2=d\sigma_+d\sigma_--d\phi^2,
\end{equation}
where $\sigma_+\equiv x^0+x^1$ and $\sigma_-\equiv x^0-x^1$.
For the solution $\phi(x)$ in the region I, $(x^0)^2-(x^1)^2>0$
the metric can be
diagonalized by using the parameters $(t,q)$, given by
\begin{equation}
t\equiv\sqrt{(x^0)^2-(x^1)^2},~~~q\equiv\phi=s\ln\frac{x^0+x^1}{x^0-x^1}.
\label{eq:parameters-ex}
\end{equation}
The detailed steps are following.
Defining first $s\ln(\sigma_+/|s|)\equiv\tilde{\sigma}_+$ and $s\ln(\sigma_-/|s|)\equiv\tilde{\sigma}_-$,
and then $\phi\equiv\tilde{\sigma}_+-\tilde{\sigma}_-$ and
$T\equiv\tilde{\sigma}_++\tilde{\sigma}_-$,
the line element can be rewritten as

\begin{equation}
\begin{array}{ccl}
ds^2&=&e^{(\tilde{\sigma}_++\tilde{\sigma}_-)/s}d\tilde{\sigma}_+d\tilde{\sigma}_--d\phi^2\\
&=&e^{T/s}\frac{1}{4}(dT^2-d\phi^2)-d\phi^2\\
&=&\frac{1}{4}e^{T/s}dT^2-(\frac{1}{4}e^{T/s}+1)d\phi^2\\
&=&dt^2-(\frac{t^2}{4s^2}+1)dq^2.
\end{array} \label{eq:exp}
\end{equation}
This induced metric represents the Robertson-Walker type spacetime in $(1+1)$ dimension with the scale factor $a(t)=\sqrt{\frac{t^2}{4s^2}+1}$.
The scale change proceeds through a time reversal symmetry.
The universe starts to shrink at $t=-\infty$ and approaches to a flat spacetime while the brane in this region becomes vertical at $t=0$. Then the universe starts to expand symmetrically in time.
Interesting is that the solution describes
a shrinking and expanding universe
connected by ``Big Bounce", where no singularities exist.
%in its geometry.
The existing singular points in the bulk coordinate seem to be originated from this particular static gauge which assumes a vacuum state to be perpendicular to the brane.
Although in higher than $(1+1)$ dimensions our solution may be just a toy model presenting an anisotropic expansion, a search for a solution having isotropic expansion in higher dimensions would be an interesting future project.

\begin{figure}[h]
  \begin{center}
\begin{tabular}{cc}
    \includegraphics[scale=0.7]{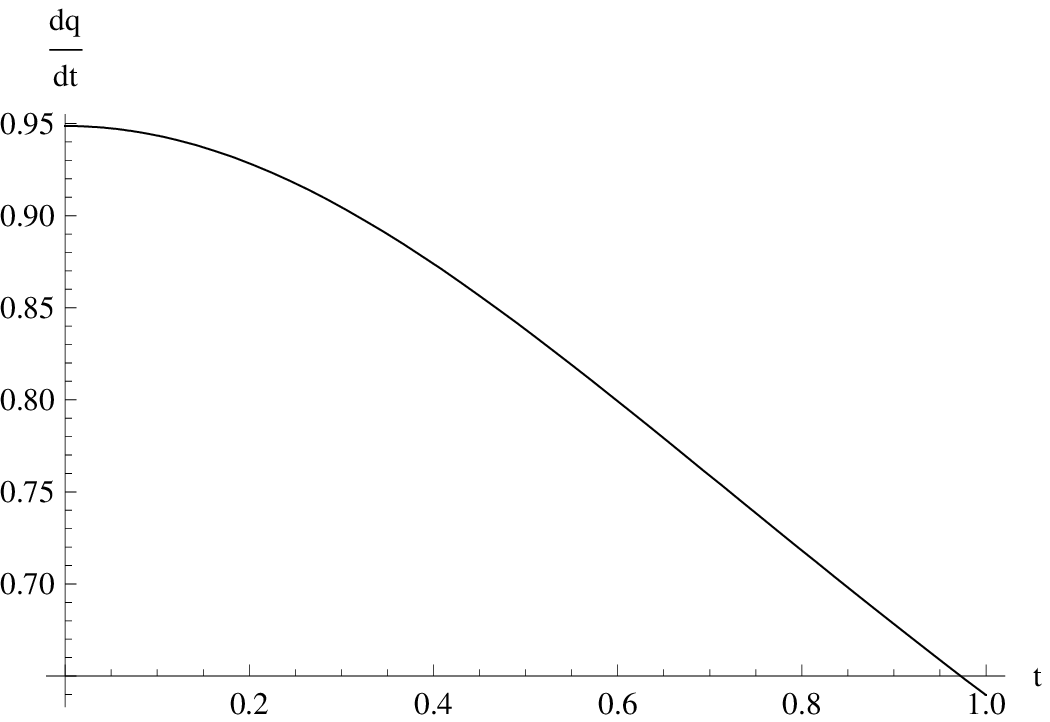} &
    \includegraphics[scale=0.7]{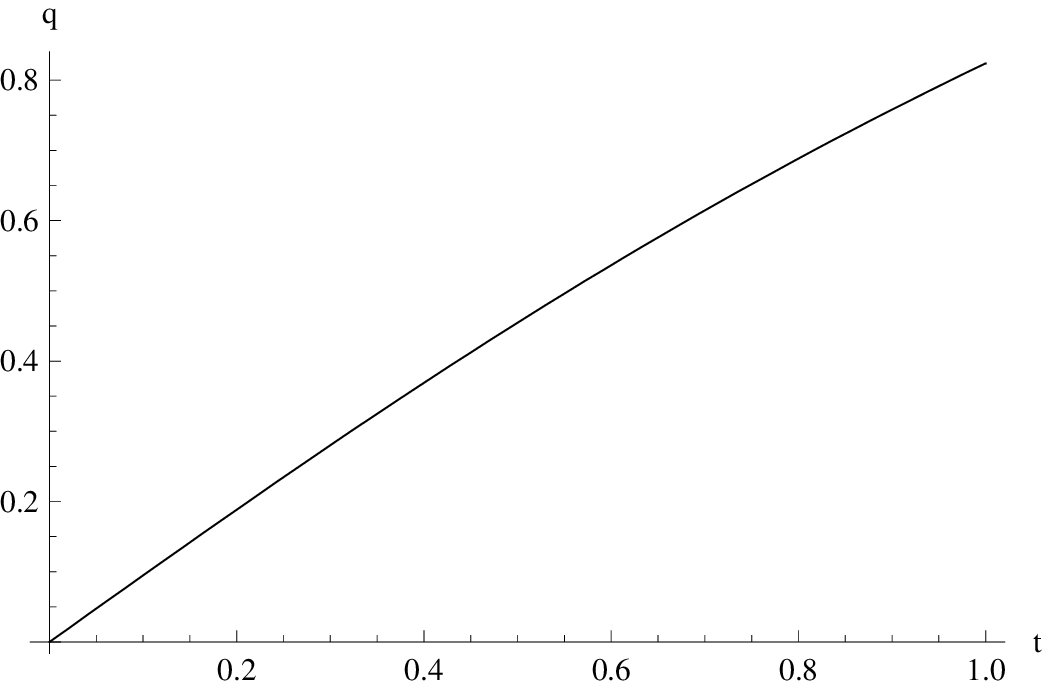} \\
\end{tabular}
    \end{center}
\caption{On the Big Bounce brane universe (a) $\frac{dq}{dt}(t)$ and (b) $q(t)$, with $p_q/m=3$, $q(0)=0$ and $s=\frac{1}{2}$.
}\label{fig:dq(t)/dt and q(t)-BigBounce}
\end{figure}

To see a classical behavior of a massive particle we need to
calculate geodesics. To this end we calculate geometric quantities here.
The Christoffel symbols
\begin{equation}
\Gamma^a_{bc}=\frac{1}{2}g^{am}(g_{mb,c}+g_{mc,b}-g_{bc,m})
\end{equation}
have the following nonzero components %of the Christoffel symbols
in the coordinates $(t,q)$:
\begin{equation}
\begin{array}{ccl}
 \Gamma^1_{10}=\frac{1}{2}g^{11}g_{11,0}=\frac{t}{t^2+4s^2}, \quad
 \Gamma^0_{11}=-\frac{1}{2}g^{00}g_{11,0}=\frac{t}{4s^2}.
\end{array}
\end{equation}
The Riemann tensor
\begin{equation}
R^a_{bcd}=\Gamma^a_{bd,c}-\Gamma^a_{bc,d}+\Gamma^a_{mc}\Gamma^m_{bd}-\Gamma^a_{md}\Gamma^m_{bc}
\end{equation}
%The only independent component of the Riemann tensor in $1+1$ dimension is
is calculated to yield
\begin{equation}
\begin{array}{ccl}
R^1_{010}&=&0-\partial_0\Gamma^1_{01}+0-\Gamma^1_{10}\Gamma^1_{01}
=-\frac{4s^2}{(t^2+4s^2)^2}.
\end{array}
\end{equation}
The Ricci scalar is found to be
%is expressed by $R^1_{010}$,
%and the Ricci scalar
\begin{equation}
R=g^{ab}R_{ab}=g^{11}g^{00}g_{11}R^1_{010}+g^{00}R^1_{010}=2g^{00}R^1_{010}
%\end{equation}
%\begin{equation}
%R
=-\frac{8s^2}{(t^2+4s^2)^2}.
\end{equation}
Now we are ready to get geodesic equations for a massive particle. They are obtained as
\begin{equation}
\left\{
\begin{array}{l}
\frac{d^2q}{d\tau^2}+2\Gamma^1_{10}\frac{dq}{d\tau}\frac{dt}{d\tau}
 = \frac{d^2q}{d\tau^2}
 +\frac{2t}{t^2+4s^2}\frac{dq}{d\tau}\frac{dt}{d\tau} =0,\\
\frac{d^2t}{d\tau^2}+\Gamma^0_{11}\left(\frac{dq}{d\tau}\right)^2
 = \frac{d^2t}{d\tau^2}+\frac{t}{4s^2}\left(\frac{dq}{d\tau}\right)^2 = 0.
\end{array}\right.
\end{equation}
These can be reduced to
\begin{equation}
\left(\frac{dv}{dt}+\frac{2t}{t^2+4s^2}v\right)w=0, \quad
\frac{1}{2}\frac{dw^2}{d\tau}+\frac{t}{4s^2}v^2=0,
\label{eq:w-ex}
\end{equation}
where $v=\frac{dq}{d\tau}$ and $w=\frac{dt}{d\tau}$.
The equation for $v(t)$ is integrated to give
\begin{equation}
v(t)=\frac{dq}{d\tau}=\frac{dq}{dt}\frac{dt}{d\tau}=\frac{v(t_0)(t_0^2+4s^2)}{t^2+4s^2}.
\label{eq:v-ex}
\end{equation}
Then $w(t)$ can be found when $v(t)$ is substituted into Eq.(\ref{eq:w-ex}),
\begin{equation}
0=\frac{1}{2}\frac{dw^2}{dt}+v^2(t_0)\frac{t}{4s^2}\left(\frac{t_0^2+4s^2}{t^2+4s^2}\right)^2 ,
\end{equation}
yielding
\begin{equation}
w(t)=\pm\sqrt{w^2(t_0)+\frac{v^2(t_0)(t_0^2+4s^2)}{4s^2}\left(\frac{t_0^2+4s^2}{t^2+4s^2}-1\right)}.
\end{equation}
The coordinate velocity $\frac{dq}{dt}$ is obtained from $v(t)$ and $w(t)$
as
\begin{equation}
\frac{dq}{dt}=\frac{v(t_0)(t_0^2+4s^2)}{t^2+4s^2}\frac{1}{w}
=\pm\frac{1}{\sqrt{\left[\frac{w^2(t_0)}{v^2(t_0)}-\frac{1}{4s^2}(t^2_0+4s^2)\right]\left(\frac{t_0^2+4s^2}{t^2+4s^2}\right)^2+\frac{t^2+4s^2}{4s^2}}}.
\end{equation}
Since only one constant is needed in solving the first order differential equation, one of $v(t_0)$ and $w(t_0)$ must be eliminated. They are related in the normalization condition for a massive particle:
\begin{equation}
\begin{array}{ccl}
V^mV_m&=&w^2g_{00}+v^2g_{11}\\
&=&\left[w^2(t_0)+\frac{v^2(t_0)(t_0^2+4s^2)}{4s^2}\left(\frac{t_0^2+4s^2}{t^2+4s^2}-1\right)\right]\times1
+\left[\frac{v(t_0)(t_0^2+4s^2)}{t^2+4s^2}\right]^2\times\frac{-(t^2+4s^2)}{4s^2}\\
&=&v^2(t_0)\left(\frac{w^2(t_0)}{v^2(t_0)}-\frac{t^2_0+4s^2}{4s^2}\right)=1.
\end{array}
\end{equation}
Thus, the coordinate velocity is simplified as
\begin{equation}
\frac{dq}{dt}
=\pm\frac{1}{\sqrt{\frac{1}{v^2(t_0)}\frac{(t^2+4s^2)^2}{(t^2_0+4s^2)^2}+\frac{t^2+4s^2}{4s^2}}}.
\end{equation}
Since the metric does not depend on $q$, the conserved quantity $\frac{dq}{d\tau}g_{11}\equiv p_q/m= vg_{11}=-v(t)(t^2+4s^2)$, where $m$ is a mass of a particle, is found. It can be also seen in Eq.(\ref{eq:v-ex}). The particle trajectory is shown in Fig.~\ref{fig:dq(t)/dt and q(t)-BigBounce}.
%%%%%%%%%%%%%%%%%%%%%%%%%%%%%%%%%%%%%%%%%%%%%%%%%%%%%
%%%%%%%%%%%%%%%%%%%%%%%%%%%%%%%%%%%%%%%%%%%%%%%%%%%%%%%

One may want to make sure that a speed of a massive particle in any frame should not exceed the speed of light. A speed of a massive particle $|\vec{v}_{bke}|$ on the brane from an observer in the bulk is expressed by
\begin{equation}
|\vec{v}_{bke}|=\sqrt{\left(\frac{dq}{dt}\right)^2+\left(\frac{dx^1}{dt}\right)^2}\left|\frac{dt}{dx^0}\right|.
\end{equation}
Here, $\frac{dx^1}{dt}$ and $\frac{dt}{dx^0}$ need to be expressed by the equation of motion. Using the transformations from Eq.(\ref{eq:parameters-ex}),
\begin{equation}
\left\{
\begin{array}{l}
x^0=t\cosh\frac{q}{2s}\\
x^1=t\sinh\frac{q}{2s},
\end{array}\right.
\end{equation}
and hence
\begin{equation}
\left\{
\begin{array}{l}
\frac{dx^0}{dt}=\frac{x^0}{t}+\frac{x^1}{2s}\frac{dq}{dt}\\
\frac{dx^1}{dt}=\frac{x^1}{t}+\frac{x^0}{2s}\frac{dq}{dt}.
\end{array}\right.
\end{equation}
The following inequality is expressed enough to be verified:
\begin{equation}
\vec{v}^2_{bke}=\frac{\left(\frac{dq}{dt}\right)^2
+\left(\frac{x^1}{t}\right)^2+\left(\frac{x^0}{2s}\right)^2\left(\frac{dq}{dt}\right)^2
+\frac{2x^0x^1}{2st}\left(\frac{dq}{dt}\right)}
{
\left(\frac{x^0}{t}\right)^2+\left(\frac{x^1}{2s}\right)^2\left(\frac{dq}{dt}\right)^2
+\frac{2x^0x^1}{2st}\left(\frac{dq}{dt}\right)}<1.
\end{equation}
This inequality is immediately reduced to the following simplified form to show $\vec{v}^2_{bke}$ is manifestly less than 1:
\begin{equation}
\frac{t^2+4s^2}{4s^2}\left(\frac{dq}{dt}\right)^2-1
=\frac{\frac{t^2+4s^2}{4s^2}}{\frac{1}{v^2(t_0)}\frac{(t^2+4s^2)^2}{(t^2_0+4s^2)^2}+\frac{(t^2+4s^2)}{4s^2}}
-1<0.
\end{equation}

%%%%%%%%%%%%%%%%%%%%%%%%%%%%%%%%%%%%%%%%%%%%%%%%%%%%%%%%%%%%%
\section{Colliding branes and reconnection}\label{sec:Colliding branes}
\begin{figure}[h]
  \begin{center}
\begin{tabular}{cc}
    \includegraphics[scale=0.7]{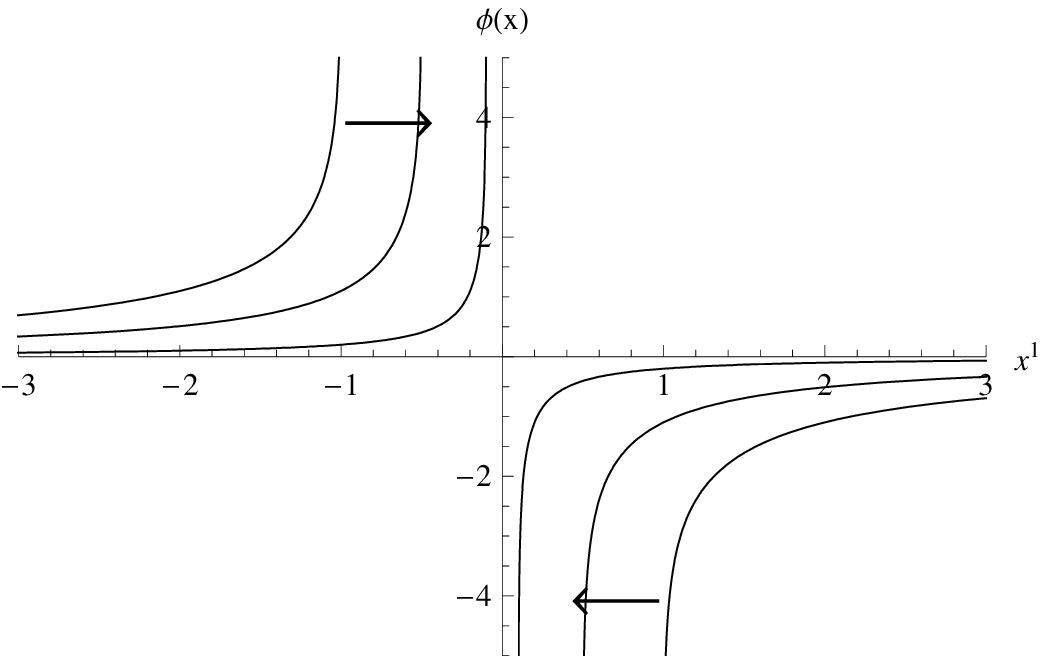} &
    \includegraphics[scale=0.7]{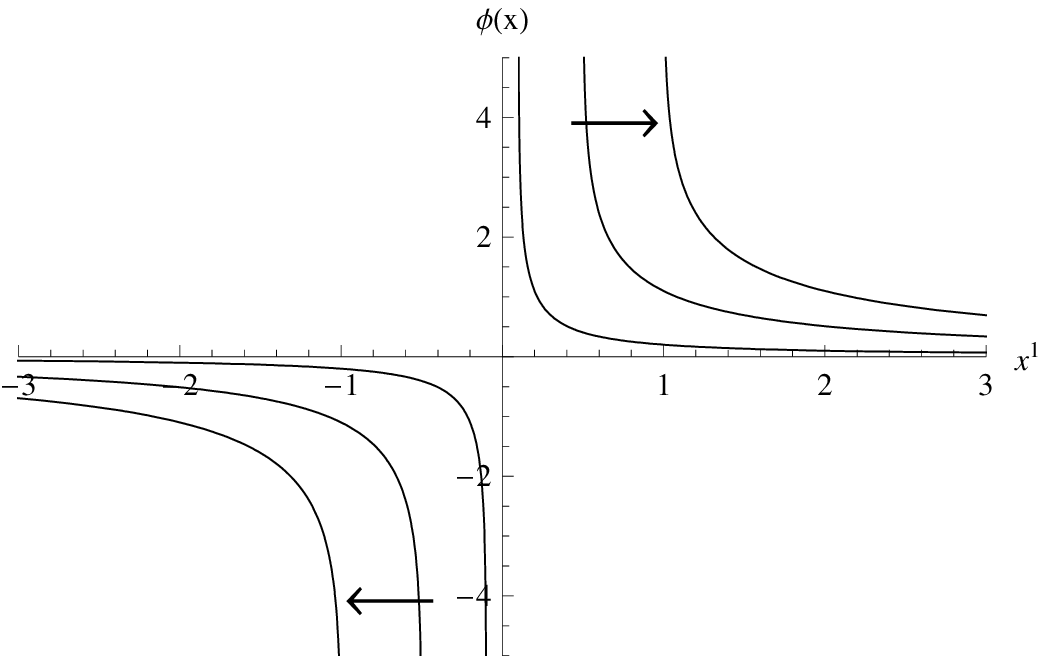} \\
   (a) & (b)
\end{tabular}
    \end{center}
\caption{The dynamics of the colliding branes. (a) The branes at $x^0=-1, -0.5, -0.1$ with $s=1$.
The branes approach each other as time goes on.
(b) The branes at $x^0=0.1, 0.5, 1$ with $s=1$.
The branes reconnect and separate each other as time goes on.
The arrows represent the directions
which the branes are moving to.
\label{fig:colliding-branes}}
\end{figure}

Let us move on to the region II, $(x^1)^2\geq(x^0)^2$.
The branes can be seen in Fig.~\ref{fig:colliding-branes}.
%%%%%%%%%%%%%%%%%%%%%%%%%%%%%
%%%%%%%%%%%%%%%%%%%%%
The metric can be
diagonalized by the parameters $(\bar{t},\bar{q})$, defined by
\begin{equation}
\bar{t}\equiv\phi=s\ln\frac{x^0+x^1}{x^1-x^0},~~~\bar{q}\equiv\sqrt{(x^1)^2-(x^0)^2} .
\label{eq:parameters-col}
\end{equation}
Changing the parameters similarly to the previous case but considering $x^1>x^0$, first define $s\ln(\sigma_+/|s|)=\xi_+$ and $s\ln(-\sigma_-/|s|)=\xi_-$ and then $\phi=\xi_+-\xi_-$ and $X=\xi_++\xi_-$, giving
\begin{equation}
\begin{array}{ccl}
ds^2&=&-e^{(\xi_++\xi_-)/s}d\xi_+d\xi_--d\phi^2\\
&=&-e^{X/s}\frac{1}{4}(dX^2-d\phi^2)-d\phi^2\\
&=&-\frac{1}{4}e^{X/s}dX^2+(\frac{1}{4}e^{X/s}-1)d\phi^2\\
&=&(\frac{\bar{q}^2}{4s^2}-1)d\bar{t}^2-d\bar{q}^2.
\end{array}
\end{equation}
Although the situation looks similar to the region I $(x^1)^2\leq(x^0)^2$,
the geometry in the region II becomes quite different due to a sign change in the logarithm. The solution for the region I is transformed to one for the region II by the exchange of the coordinates,
\begin{equation}
x^0\Leftrightarrow x^1.
\end{equation}
However, since the line element $(dx^0)^2-(dx^1)^2$ stays unchanged, the geometries in the regions I and II are not symmetrical in the exchange of the coordinate in the solution.
Nevertheless, it is worthwhile to recognize the diagonalized metric here can be just obtained from one (\ref{eq:exp}) for the Big Bounce brane universe
%found in the last section
by the transformation
\begin{equation}
t\rightarrow \pm i\bar{q},~~~q\rightarrow \bar{t}. \label{eq:transform I to II}
\end{equation}
Furthermore, a change of sign of the determinant of the metric occurs at $\bar{q}=\pm\sqrt{(x^1)^2-(x^0)^2}=\pm2s$.
This suggests that this region of the solution should be divided into
the Minkowskian ($|\bar{q}|=\sqrt{(x^1)^2-(x^0)^2}>2|s|$) and the Euclidean spaces ($|\bar{q}|=\sqrt{(x^1)^2-(x^0)^2}<2|s|$).
Although many references discuss
a model having the Minkowski and the Euclidean regions both \cite{Gibbons:2004dz}, its legitimacy needs to be clear. As long as the entire solution is not abandoned, the Euclidean region must be kept because of a continuation of the energy flow.

\begin{figure}[h]
  \begin{center}
\begin{tabular}{cc}
    \includegraphics[scale=0.7]{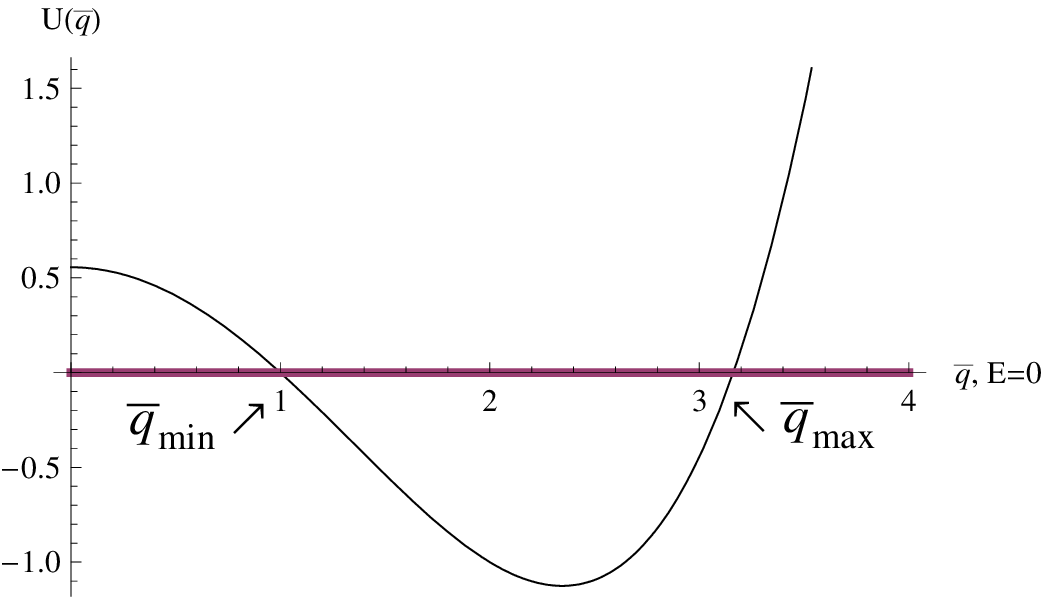} &
    \includegraphics[scale=0.7]{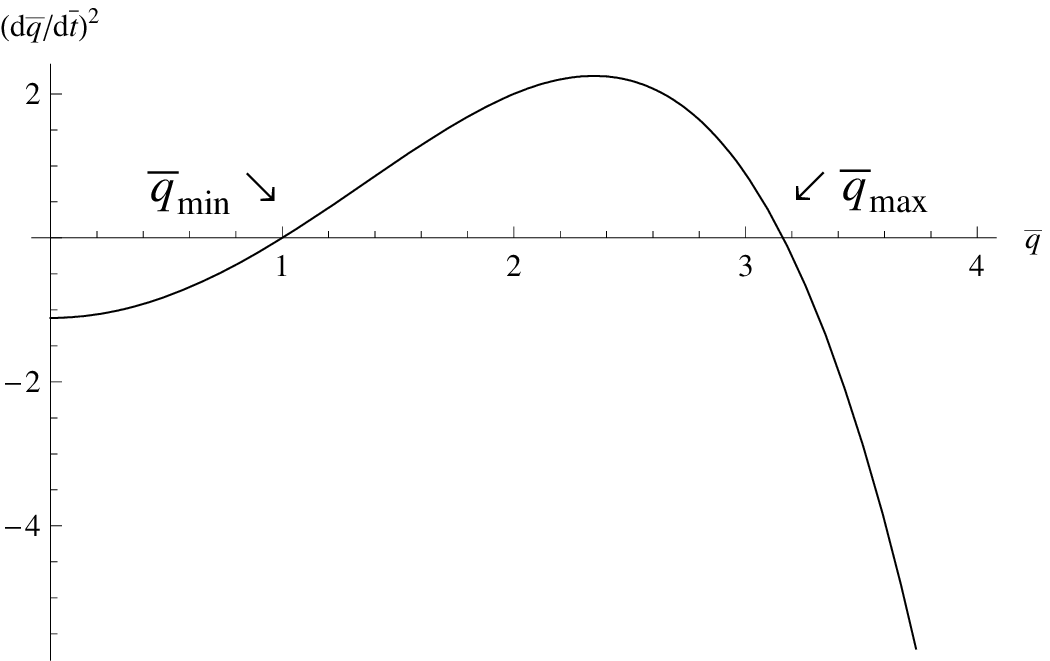} \\
    (a) & (b)
\end{tabular}
    \end{center}
\caption{On the colliding brane (a) $U(\bar{q})$ and
(b)$\frac{d\bar{q}}{d\bar{t}}$ vs. $\bar{q}$, with $\bar{p}_{\bar{t}}/m=3$, $s=\frac{1}{2}$ and $m=1$.}
\label{fig:U(q) and dq/dt(q)-col}
\end{figure}

For a moment we do not have to restrict ourselves to only
the Minkowski region until we get the final result.
Following the same routines as in the last section, the necessary Christoffel symbols, curvature tensor and scalar can be found,
\begin{equation}
\begin{array}{ccl}
&& \Gamma^0_{01}=\frac{1}{2}g^{00}g_{00,1}=\frac{\bar{q}}{\bar{q}^2-4s^2}, \quad
   \Gamma^1_{00}=-\frac{1}{2}g^{11}g_{00,1}=\frac{\bar{q}}{4s^2} ,\\
%\end{array}
%\end{equation}
%\begin{equation}
%\begin{array}{ccl}
&& R^1_{010}=\partial_1\Gamma^1_{00}-0+0-\Gamma^1_{00}\Gamma^0_{01}=-\frac{1}{\bar{q}^2-4s^2},\\
&& R=2g^{00}R^1_{010}=\frac{-8s^2}{(\bar{q}^2-4s^2)^2}.
\end{array}
\end{equation}
The geodesic equations for a massive particle read
\begin{equation}
\left\{
\begin{array}{l}
\frac{d^2\bar{q}}{d\tau^2}+\Gamma^1_{00}\left(\frac{d\bar{t}}{d\tau}\right)^2
= \frac{d^2\bar{q}}{d\tau^2}+\frac{\bar{q}}{4s^2}\left(\frac{d\bar{t}}{d\tau}\right)^2=0,\\
 \frac{d^2\bar{t}}{d\tau^2}+2\Gamma^0_{01}\frac{d\bar{t}}{d\tau}\frac{d\bar{q}}{d\tau}
= \frac{d^2\bar{t}}{d\tau^2}+\frac{2\bar{q}}{\bar{q}^2-4s^2}\frac{d\bar{t}}{d\tau}\frac{d\bar{q}}{d\tau}=0.
\end{array}\right.
\end{equation}
It can be realized that the Ricci scalar and the geodesics in the region I are transformed to those in the region II
by Eq.(\ref{eq:transform I to II}), ($t\rightarrow \pm i\bar{q}$, $q\rightarrow \bar{t}$).
The geodesic equations are reduced with a familiar form as before,
\begin{equation}
\frac{1}{2}\frac{d\bar{v}^2}{d\bar{q}}+\frac{\bar{q}}{4s^2}\bar{w}^2=0, \quad
\left(\frac{d\bar{w}}{d\bar{q}}+\frac{2\bar{q}}{\bar{q}^2-4s^2}\bar{w}\right)\bar{v}=0,
\end{equation}
where $\bar{v}=\frac{d\bar{q}}{d\tau}$ and $\bar{w}=\frac{d\bar{t}}{d\tau}$.
The equation for $w(\bar{q})$ is integrated giving
\begin{equation}
\bar{w}(\bar{q})=\bar{w}(\bar{q}_0)\frac{\bar{q}_0^2-4s^2}{\bar{q}^2-4s^2} .
\end{equation}
Plugging it into the equation for $\bar{v}(\bar{q})$, we have
\begin{equation}
0=\frac{1}{2}\frac{d\bar{v}^2}{d\bar{q}}+\frac{\bar{q}}{4s^2}\bar{w}^2(\bar{q}_0)\frac{(\bar{q}_0^2-4s^2)^2}{(\bar{q}^2-4s^2)^2} ,
\end{equation}
which is integrated to
\begin{equation}
\begin{array}{ccl}
\bar{v}^2(\bar{q})&=&\frac{\bar{w}^2(\bar{q}_0)}{4s^2}(\bar{q}_0^2-4s^2)^2\left[\frac{1}{\bar{q}^2-4s^2}-\frac{1}{\bar{q}^2_0-4s^2}\right]+v^2(\bar{q}_0)\\
&=&\frac{\bar{w}^2(\bar{q}_0)}{4s^2}(\bar{q}^2_0-\bar{q}^2)\frac{\bar{q}_0^2-4s^2}{\bar{q}^2-4s^2}+v^2(\bar{q}_0),
\end{array}
\end{equation}
and therefore we obtain
\begin{equation}
\bar{v}(\bar{q})=\pm\sqrt{\frac{\bar{w}^2(\bar{q}_0)}{4s^2}\frac{(\bar{q}_0^2-4s^2)^2}{\bar{q}^2-4s^2}-\frac{\bar{w}^2(\bar{q}_0)}{4s^2}(\bar{q}_0^2-4s^2)+\bar{v}^2(\bar{q}_0)}.
\label{eq:v-col}
\end{equation}
The coordinate velocity can be calculated with $\bar{v}(\bar{q})$ and $\bar{w}(\bar{q})$,
\begin{equation}
\frac{d\bar{q}}{d\bar{t}}=\frac{1}{\bar{w}(\bar{q}_0)}\frac{\bar{q}^2-4s^2}{\bar{q}^2_0-4s^2}\bar{v}(\bar{q})
=\pm\sqrt{\left(\frac{\bar{v}^2(\bar{q}_0)}{\bar{w}^2(\bar{q}_0)}-\frac{\bar{q}_0^2-4s^2}{4s^2}\right)\left(\frac{\bar{q}^2-4s^2}{\bar{q}^2_0-4s^2}\right)^2+\frac{\bar{q}^2-4s^2}{4s^2}} .
\label{eq:dq/dt-col}
\end{equation}
When the normalization condition for the velocity vector
\begin{equation}
\begin{array}{ccl}
V^mV_m&=&\bar{w}^2g_{00}+\bar{v}^2g_{11}\\
&=&(\frac{\bar{q}^2}{4s^2}-1)\bar{w}^2(\bar{q}_0)\frac{(\bar{q}_0^2-4s^2)^2}{(\bar{q}^2-4s^2)^2}
-\frac{\bar{w}^2(\bar{q}_0)}{4s^2}(\bar{q}_0^2-4s^2)^2\left[\frac{1}{\bar{q}^2-4s^2}-\frac{1}{\bar{q}^2_0-4s^2}\right]-\bar{v}^2(\bar{q}_0)\\
&=&\frac{\bar{w}^2(\bar{q}_0)}{4s^2}(\bar{q}_0^2-4s^2)-\bar{v}^2(\bar{q}_0)=1
\end{array}\label{eq:normalization-col}
\end{equation}
is applied to Eq.(\ref{eq:dq/dt-col}), the coordinate velocity is reduced to
\begin{equation}
\frac{d\bar{q}}{d\bar{t}}=\pm\sqrt{\frac{\bar{q}^2-4s^2}{4s^2}-\frac{1}{\bar{w}^2(\bar{q}_0)}\left(\frac{\bar{q}^2-4s^2}{\bar{q}^2_0-4s^2}\right)^2}.
\end{equation}
If $\bar{q}^2<4s^2$, which corresponds to the Euclidean region, $x^0<x^1<x^1_c=\sqrt{(x^0)^2+4s^2}$, the coordinate velocity $\frac{d\bar{q}}{dt}$
becomes imaginary, which is not allowed in classical mechanics.
However, one can hardly imagine that the Euclidean region can be ignored in quantum mechanics when viewed as a motion of a non-relativistic particle with $E=0$ under the effective potential energy $U(\bar{q})$ given by
\begin{equation}
U(\bar{q})=-\frac{m}{2}\left[\frac{\bar{q}^2-4s^2}{4s^2}-\frac{1}{\bar{w}^2(\bar{q}_0)}\left(\frac{\bar{q}^2-4s^2}{\bar{q}^2_0-4s^2}\right)^2\right].
\end{equation}
The potential $U(q)$ is visualized in Fig.~\ref{fig:U(q) and dq/dt(q)-col}.
%%%%%%%%%%%%%%%%%%%%%%%%%%%%%%%%%%%%%%%%%v2cols05.eps
%%%%%%%%%%%%%%%%%%%%%%%%%%%%%%%%%%%%%%%%%
Apparently, the allowed region ($U<0$) is
\begin{equation}
4s^2<q^2\leq\frac{\bar{w}^2(\bar{q}_0)(\bar{q}^2_0-4s^2)^2}{4s^2}+4s^2.
\end{equation}
Since this inequality also holds at $\bar{q}=\bar{q}_0$, the following relation is induced
\begin{equation}
\bar{w}^2(\bar{q}_0)(\bar{q}_0^2-4s^2)\geq4s^2.
\end{equation}
However, this is nothing but the normalization condition
Eq.~(\ref{eq:normalization-col}). It can be reexpressed in the conserved quantity $\frac{dt}{d\tau}g_{00}\equiv \bar{p}_{\bar{t}}/m= \frac{w(q)(q^2-4s^2)}{4s^2}\geq 1$.
The two boundary points $(\bar{q}_{min},\bar{q}_{max})=(2s,2s\sqrt{(\bar{p}_{\bar{t}}/m)^2+1})$ and
the minimum potential point
$\bar{q}_c=2s\sqrt{(\bar{p}_{\bar{t}}/m)^2/2+1}$ can be written in terms of $\bar{p}_{\bar{t}}$, $m$ and $s$.
%%%%%%%%%%%%%%%%%%%%%%%%%%%%%%%%%%%%%%%

\begin{figure}[h]
  \begin{center}
\begin{tabular}{cc}
    \includegraphics[scale=0.7]{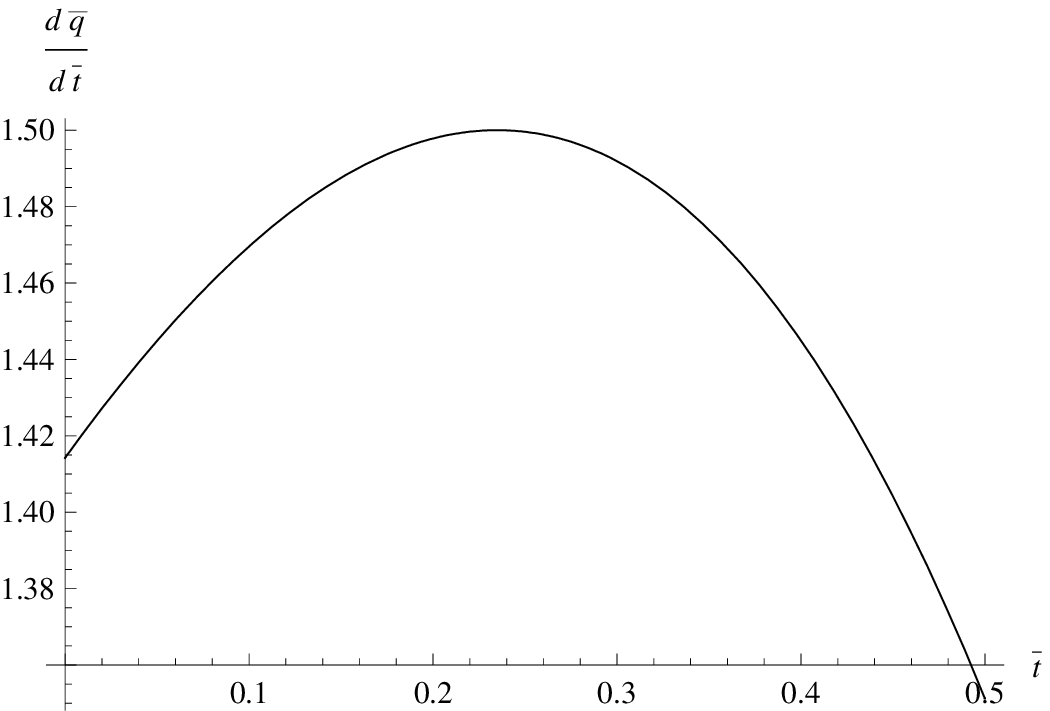} &
    \includegraphics[scale=0.7]{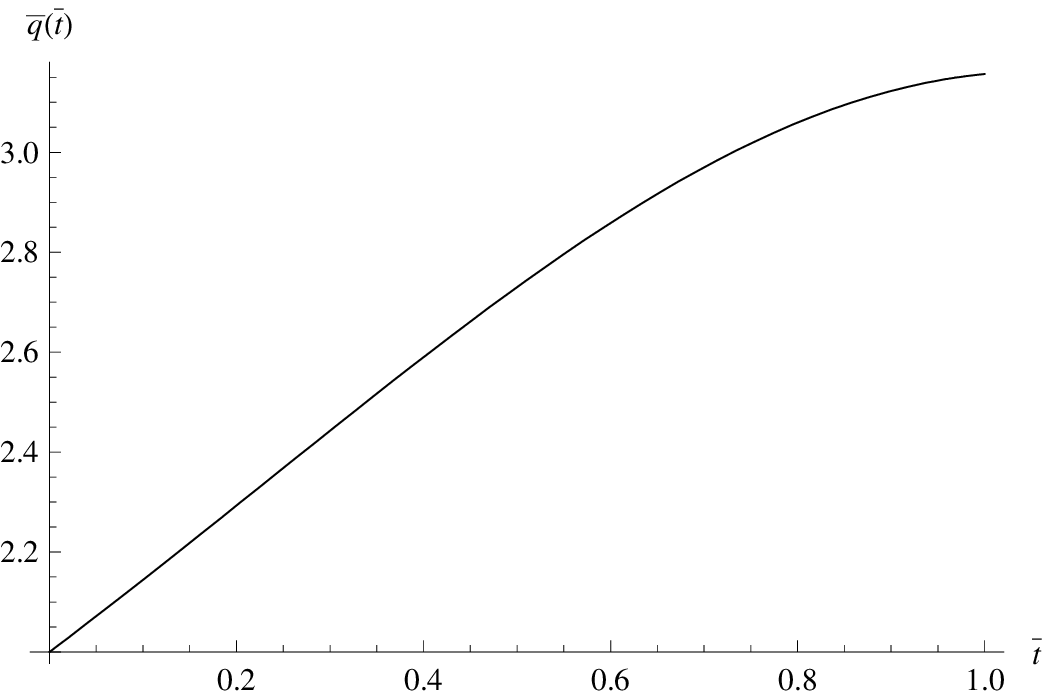} \\
    (a) & (b)
\end{tabular}
    \end{center}
\caption{On the colliding brane
(a) $\frac{d\bar{q}}{d\bar{t}}$ vs. $\bar{t}$ and
(b) $\bar{q}(\bar{t})$, with $\bar{p}_{\bar{t}}/m=3$ and $s=\frac{1}{2}$.}
\label{fig:dq(t)/dt and q(t)-col}
\end{figure}

A speed of a massive particle on the brane from an observer in the bulk is
\begin{equation}
\vec{v}^2_{bkc}=\left[\left(\frac{d\bar{q}}{d\bar{t}}\right)^2
+\left(\frac{dx^1}{d\bar{t}}\right)^2\right]\left(\frac{d\bar{t}}{dx^0}\right)^2.
\end{equation}
Using the transformations Eq.~(\ref{eq:parameters-col})
\begin{equation}
\left\{
\begin{array}{l}
x^0=\bar{q}\sinh\frac{\bar{t}}{2s},\\
x^1=\bar{q}\cosh\frac{\bar{t}}{2s},
\end{array}\right.
\end{equation}
and hence
\begin{equation}
\left\{
\begin{array}{l}
\frac{dx^0}{d\bar{t}}=\frac{d\bar{q}}{dt}\frac{x^0}{\bar{q}}+\frac{x^1}{2s},\\
\frac{dx^1}{d\bar{t}}=\frac{d\bar{q}}{dt}\frac{x^1}{\bar{q}}+\frac{x^0}{2s},
\end{array}\right.
\end{equation}
we obtain
\begin{equation}
\vec{v}^2_{bkc}=\frac{1+
\left(\frac{d\bar{q}}{d\bar{t}}\right)^2\left(\frac{x^1}{\bar{q}}\right)^2+\left(\frac{x^0}{2s}\right)^2
+\frac{2x^0x^1}{2s\bar{q}}\left(\frac{d\bar{q}}{d\bar{t}}\right)}
{
\left(\frac{d\bar{q}}{d\bar{t}}\right)^2\left(\frac{x^0}{\bar{q}}\right)^2+\left(\frac{x^1}{2s}\right)^2
+\frac{2x^0x^1}{2s\bar{q}}\left(\frac{d\bar{q}}{d\bar{t}}\right)}<1
\end{equation}
is required for a massive particle.
This inequality becomes
\begin{equation}
1+\left(\frac{d\bar{q}}{d\bar{t}}\right)^2-\frac{\bar{q}^2-4s^2}{4s^2}
=-\frac{1}{w^2(x_0)}\left(\frac{\bar{q}^2-4s^2}{\bar{q}^2_0-4s^2}\right)^2<0,
\end{equation}
indeed showing $\vec{v}^2_{bkc}$ is manifestly less than 1 even in the Euclidean region.
We plot the coordinate velocity $\frac{d\bar{q}}{d\bar{t}}$ versus $\bar{t}$ and the corresponding potential in Fig.~\ref{fig:dq(t)/dt and q(t)-col}.
%%%%%%%%%%%%%%%%%%%%%%%%%%%%%%%%%%%%%%%%%v2cols05.eps
%%%%%%%%%%%%%%%%%%%%%%%%%%%%%%%%%%%%%%%%%

%%%%%%%%%%%%%%%%%%%%%%%%%%%%%%%%%%%%%%%%%%
%\newpage
\section{Conclusion and Discussion}\label{sec:Conclusion and Discussion}
A wave solution of the Nambu-Goto action with codimension one in the static
gauge has been studied. The solution represents two-wave scattering
as in Fig.~\ref{fig:log-sol}.
An interesting feature is that the single solution is naturally
decomposed into the regions having two different geometries by its singularity.
% and a signature change.
That is, the whole brane consists of the central brane and the outside
branes, representing the Robertson-Walker type universe and the colliding branes,
as seen in Fig.~\ref{fig:brane_universe} and Fig.~\ref{fig:colliding-branes}, respectively.
These two geometries are transformed by a simple coordinate exchange, but the dynamics of particles there appears quite different in the end.
The former describes the expanding and shrinking universe
connected by the ``Big Bounce" which does not have ``Big Bang" singularities in its geometry.
The latter represents that two branes collide and reconnect each other.
Specially, it has two different geometries through a signature change
in the metric.
The detailed classical dynamics of a massive particle on the branes has been provided.
In the end, it has been verified that a speed of a massive particle on the brane cannot exceed the speed of light. As given in Appx.~\ref{appx:Energy of brane}, the normalized energy of brane in the bulk turns out to be infinite due to the Big Bounce universe and the Euclidean parts of the colliding branes.
Only the Minkowskian parts of colliding branes have a finite energy.

%%%%%%%%%%%%%%%
\medskip
Before closing this paper,
several discussions are addressed here.
The colliding branes consist of the Euclidean and
Minkowski regions as studied previously in \cite{Gibbons:2004dz}.
Dealing with the Euclidean region is subtle, since it makes the action and the energy density imaginary but it is a part of an analytically continuous solution which leads a smooth energy flow.
Hopefully, two infinities from the Big Bounce universe and the Euclidean brane could be accidentally canceled in a certain circumstance if the definition of energy is extended to a complex space.

The solutions found in this paper are not the most general
for waves propagating into two directions,
because we have first assumed that solutions satisfy the Klein-Gordon equation $\partial^2 \phi=0$
in the form of $\phi(x) = f(k\cdot x) + g(p\cdot x)$.
When two waves are made simultaneously
in well separated regions of a brane,
each of them keeps the same form (\ref{eq:single wave}) without interference,
and the solution can be approximately written as
a sum of two waves if they are not much overlapped.
However, once those waves get close to each other
such an approximation is no longer valid.
Therefore,
we have to solve a generic case of two colliding waves
with the asymptotic boundary conditions
of two well separated waves.
This should be solved without any assumptions
by solving Eq.(\ref{eq:EOM}) directly,
in which non-trivial cancelation between
contributions from the first and second terms may occur.

We have studied the case of a brane of codimension one in this paper,
corresponding to a domain wall.
Extension to higher codimensional case remains as a future problem.
Especially for the case of codimension two,
it will describe waves on (cosmic) strings.
Several solutions to string equations of motion
were extensively studied in the literature,
in particular, on the relation with the rigidity of strings
\cite{Curtright:1986vg}.
This case should be pursued further
which will be also important in study of
cosmic strings in cosmology \cite{Garfinkle:1990jq,Siemens:2001dx}.

Fundamental strings (or branes) ending on a brane can be realized as
classical solutions (or solitons)
in the effective field theory
(typically the Nambu-Goto or the Dirac-Born-Infeld action)
on the host brane \cite{Townsend:1999hi}.
This point of view is well established for static configurations
of bound states of strings and branes.
Endpoints of the fundamental strings are in fact singular
spikes in the effective theory of the host brane,
which are called BIons \cite{Callan:1997kz}.
In our solution, the two spikes are moving at the speed of light
as in Fig.~\ref{fig:log-sol}.
They may realize moving branes ending on a
host $p$-brane from its both sides.

The Big Bounce brane solution found in this paper
is isotropic only in $(1+1)$ dimensions.
Although in higher than $(1+1)$ dimension our solution
gives an anisotropic expansion,
a search for a solution having an isotropic expansion
in higher dimensions would be an interesting future project.
On the other hand the colliding brane solution
may give an interesting model for the brane world scenario.
This may be applied to the ekpyrotic universe scenario
of colliding branes \cite{Khoury:2001wf}.

We have studied classical dynamics of particular branes
and a massive particle on those branes.
One of future researches can be directed to quantization problems.
First, a massive particle motion can be quantized
in a usual manner of the first quantization.
Second, massless or massive fields localized on
the brane can be considered and can be quantized.
In particular, it is interesting to see
if there is a particle creation or annihilation for
the second quantized fields in curved space
\cite{Birell-Davis}
induced on the brane, especially for the case of
the Big Bounce universe solution.
Third, the quantization of the brane oscillation $\phi(x)$ itself
has been studied for example in \cite{Lee:2007hp}.
This can be applied to an oscillation around
a non-trivial background as found in this paper.

Finally, the brane oscillation field $\phi(x)$ becomes
a brane vector if coupled to a bulk gravity \cite{Clark:2006pd}.
Phenomenological consequences of
its coupling to the Standard Model
fields localized on the brane have been studied \cite{Clark:2007wj}.
It is an interesting direction to investigate what happens for oscillations
from particular backgrounds such as a solution found in this paper.

\appendix
%%%%%%%%%%%%%%%%%%%%%%%%%%%%%%%%%%%%%%%%%%
\section{Energy of the brane}\label{appx:Energy of brane}
\begin{figure}[h]
  \begin{center}
    \includegraphics[scale=1]{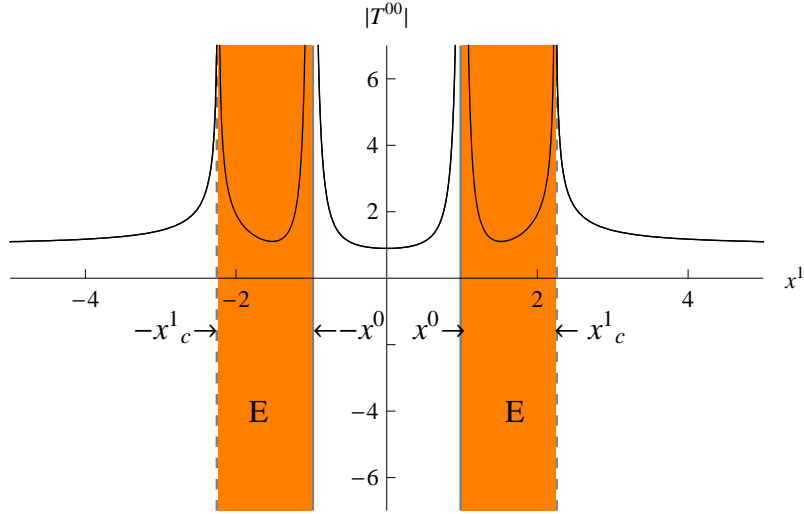}
\end{center}
\caption{The energy density is plotted with $s=1$ at $x^0=1$.
The energy density is purely imaginary in the Euclidean region,
$|x^0|<|x^1|<|x^1_c|$, which are shaded and denoted by ``E".}
\label{fig:T00}
\end{figure}

The purpose of this Appendix is to calculate the energy density and the energy in the bulk frame to see whether the energy in each of the regions is finite or not.
The energy-momentum tensor of the Nambu-Goto action
(\ref{eq:Nambu-Goto action})
can be obtained by translational invariance of the action, to yield
\begin{equation}
\begin{array}{ccl}
T^{mn}&=&\frac{\partial\mathscr{L}}{\partial(\partial_m\phi)}\partial^n\phi-\mathscr{L}\eta^{mn}\\
&=&\sigma\frac{\partial^m\phi\partial^n\phi}{\sqrt{1-(\partial\phi)^2}}+\sigma\sqrt{1-(\partial\phi)^2}\eta^{mn}\\
&=&\sigma g^{mn}\sqrt{1-(\partial\phi)^2}.
\end{array}
\end{equation}
Note that the Euclidean energy density is purely imaginary
and the Minkowskian energy density is real no matter what solution is used
\cite{Gibbons:2004dz}.
The energy density for the present solution is
\begin{equation}
\begin{array}{ccl}
T^{00}&=&\frac{\sigma}{\sqrt{1+\frac{4s^2}{(x^0)^2-(x^1)^2}}}[g+(\partial_0\phi)^2]\\
&=&\frac{\sigma}{\sqrt{1+\frac{4s^2}{(x^0)^2-(x^1)^2}}}[1-(\partial_1\phi)^2]\\
&=&\frac{\sigma}{\sqrt{1+\frac{4s^2}{(x^0)^2-(x^1)^2}}}\{1+\frac{4s^2(x^0)^2}{[(x^0)^2-(x^1)^2]^2}\}\\
&=&\sigma\sqrt{\frac{(x^0)^2-(x^1)^2}{(x_c)^2-(x^1)^2}}
+\sigma4s^2\sqrt{\frac{(x^0)^2-(x^1)^2}{(x^1_c)^2-(x^1)^2}}\frac{(x^0)^2}{[(x^0)^2-(x^1)^2]^2}.
\end{array}\label{eq:T00}
\end{equation}
See Fig.~\ref{fig:T00} for a plot of the energy $|T^{00}|$.
%%%%%%%%%%%%%%%%%%%%%%%%%%%%%%%%
%%%%%%%%%%%%%%%%%%%%%%%%%%%%%%%%
The energy densities can be written for the three separate regions
as follows:\\
1) Region I [$(x^0)^2>(x^1)^2$]:
\begin{equation}
T^{00}=\sigma\sqrt{\frac{(x^0)^2-(x^1)^2}{(x^1_c)^2-(x^1)^2}}
+\sigma4s^2(x^0)^2\frac{1}{\sqrt{(x^1_c)^2-(x^1)^2}}\frac{1}{[(x^0)^2-(x^1)^2]^{3/2}},
\label{eq:T00-I}
\end{equation}
2) Region II-E [$(x^0)^2<(x^1)^2<(x^1_c)^2$]:
\begin{equation}
T^{00}=i\sigma\sqrt{\frac{(x^1)^2-(x^0)^2}{(x^1_c)^2-(x^1)^2}}
+i\sigma4s^2(x^0)^2\frac{1}{\sqrt{(x^1_c)^2-(x^1)^2}}\frac{1}{[(x^1)^2-(x^0)^2]^{3/2}},
\label{eq:T00-II-E}
\end{equation}
3) Region II-M [$(x^1_c)^2<(x^1)^2$]:
\begin{equation}
T^{00}=\sigma\sqrt{\frac{(x^1)^2-(x^0)^2}{(x^1)^2-(x^1_c)^2}}
+\sigma4s^2(x^0)^2\frac{1}{\sqrt{(x^1)^2-(x^1_c)^2}}\frac{1}{[(x^1)^2-(x^0)^2]^{3/2}},
 \label{eq:T00-II-M}
\end{equation}
where $(x^1_c)^2=(x^0)^2+4s^2$ as defined in Eq.~(\ref{eq:Euclidean}).
Before we get the total energy
by integrating $T^{00}$ with respect to $x^1$,
we need to define the ``normalized'' energy by
subtracting the ``zero point energy'', which is the energy at $\phi=0$ from the original energy, i.e.,
\begin{equation}
 E_{\rm norm}=\int dx^1(T^{00}-\sigma).\label{eq:NormEnergy}
\end{equation}
Because $T^{00}$ is an even function in $x^1$ we can only consider the region $x^1>0$ when we see if energy is finite.
First, the following relations tell us that the first term in Eq.~(\ref{eq:T00-I}) gives a finite contribution while the second an infinite contribution;
\begin{equation}
\begin{array}{ccl}
&&\int^{x^0}_0dx^1\sqrt{\frac{(x^0)^2-(x^1)^2}{(x^1_c)^2-(x^1)^2}}<\int^{x^0}_0dx^1=\mbox{finite},\\
&&
\begin{array}{ccl}
\int^{x^0}_0dx^1\frac{1}{\sqrt{(x^1_c)^2-(x^1)^2}}\frac{1}{[(x^0)^2-(x^1)^2]^{3/2}}
&>&\frac{1}{x^1_c}\int^{x^0}_0\frac{dx^1}{[(x^0)^2-(x^1)^2]^{3/2}}\\
&=&\frac{1}{x^1_c}\frac{1}{(x^0)^2}\int^{\pi/2}_0\sec^2\theta d\theta\\
&=&\frac{1}{x^1_c}\frac{1}{(x^0)^2}\tan\theta|^{\pi/2}_0=\infty.
\end{array}
\end{array}
\end{equation}
On the other hand, since the zero point energy has the finite contribution in a finite-size region, the energy in the region I diverges.

Similarly, using the following relations the energy
in the Euclidean region in
the region II is also infinite in an imaginary direction;
\begin{equation}
\begin{array}{ccl}
&&
\begin{array}{ccl}
\int^{x^1_c}_{x^0}dx^1\sqrt{\frac{(x^1)^2-(x^0)^2}{(x^1_c)^2-(x^1)^2}}
&<&\sqrt{(x^1_c)^2-(x^0)^2}\int^{x^1_c}_{x^0}dx^1\frac{1}{\sqrt{(x^1_c)^2-(x^1)^2}}\\
&=&\sqrt{(x^1_c)^2-(x^0)^2}\int^{\pi/2}_{\arcsin\frac{x^0}{x^1_c}}d\theta
=\mbox{finite},
\end{array}\\
&&\begin{array}{ccl}
\int^{x^1_c}_{x^0}\frac{dx^1}{\sqrt{(x^1_c)^2-(x^1)^2}[(x^1)^2-(x^0)^2]^{3/2}}
&=&\int^{\pi/2}_{\arcsin\frac{x^0}{x^1_c}}\frac{d\theta}{[(x^1_c)^2\sin^2\theta-(x^0)^2]^{3/2}}\\
&=&\frac{1}{(x^1_c)^2}\int^{\pi/2}_{\arccos\frac{2s}{x^1_c}}\frac{d\theta}{[4s^2/(x^1_c)^2-\cos^2\theta]^{3/2}}\\
&>&\frac{1}{(x^1_c)^2}\int^{\pi/2}_{\arccos\frac{2s}{x^1_c}}\frac{\cos\theta d\theta}{[4s^2/(x^1_c)^2-\cos^2\theta]^{3/2}}\\
&=&\frac{1}{2(x^1_c)^2}\frac{-2}{[4s^2/(x^1_c)^2-\cos^2\theta]^{1/2}}|^{\pi/2}_{\arccos\frac{2s}{x^1_c}}\\
&=&\infty.
\end{array}
\end{array}
\end{equation}
The Minkowskian part in the region II has an finite energy by substraction of the zero point energy. Since only the first term in Eq.(\ref{eq:T00-II-M}) gives an infinite contribution by the following relations
\begin{equation}
\begin{array}{ccl}
&&\int^{\infty}_{x^1_c}dx^1\sqrt{\frac{(x^1)^2-(x^0)^2}{(x^1)^2-(x^1_c)^2}}>\int^{\infty}_{x^1_c}dx^1=\infty,\\
&&\begin{array}{ccl}
\int^{\infty}_{x^1_c}\frac{dx^1}{\sqrt{(x^1)^2-(x^1_c)}[(x^1)^2-(x^0)^2]^{3/2}}
&=&\int^{\infty}_{0}\frac{d\theta}{[(x^1_c)^2(1+\sinh^2\theta)-(x^0)^2]^{3/2}}\\
&=&\frac{1}{(x^1_c)^3}\int^{\infty}_{0}\frac{d\theta}{[\sinh^2\theta+4s^2/(x^1_c)^2]^{3/2}}\\
&<&\frac{1}{(x^1_c)^3}\int^{\infty}_{0}\frac{2\cosh\theta d\theta}{[\sinh^2\theta+4s^2/(x^1_c)^2]^{3/2}}\\
&=&\frac{1}{(x^1_c)^3}\frac{-2}{[\sinh^2\theta+4s^2/(x^1_c)^2]^{1/2}}|^\infty_0\\
&=&\mbox{finite},
\end{array}
\end{array}
\end{equation}
we concentrate on the following integral from $E_{\rm norm}$,
\begin{equation}
\begin{array}{ccl}
0<\int^{\infty}_{x^1_c}dx^1\left(\sqrt{\frac{(x^1)^2-(x^0)^2}{(x^1)^2-(x^1_c)^2}}-1\right)
&<&\int^{\infty}_{x^1_c}dx^1\left(\frac{x^1}{\sqrt{(x^1)^2-(x^1_c)^2}}-1\right)\\
&=&(\sqrt{(x^1)^2-(x^1_c)^2}-x^1)|^\infty_{x^1_c}\\
&=&x^1_c-\lim\limits_{x^1\to\infty}[x^1-\sqrt{(x^1)^2-(x^1_c)^2}].
\end{array}\label{eq:inequality-IIM2}
\end{equation}
Using the inequality $\sqrt{a^2-b^2}<|a|-|b|$ when $|a|<|b|$, $\sqrt{(x^1)^2-(x^1_c)^2}>x^1-x^1_c$ and thus $x^1-\sqrt{(x^1)^2-(x^1_c)^2}<x^1_c$.
Hence,
\begin{equation}
0<\int^{\infty}_{x^1_c}dx^1\left(\sqrt{\frac{(x^1)^2-(x^0)^2}{(x^1)^2-(x^1_c)^2}}-1\right)<x^1_c .
\end{equation}
Therefore, the energy of the Minkowskian part of the region II is finite.

The total energy is still infinite due to
the region I and the Euclidean region in the region II.
The divergence of the imaginary energy in the Euclidean region may give the possibility that the magnitude of the total energy defined in complex space could be finite, if the energy concept is extended to a complex space.

%%%%%%%%%%%%%%%%%%%%%%%%%%%%%%%%%%%%%%%%%%%
\section*{Acknowledgments}
We would like to thank T.~E.~Clark and Martin Kruczenski
for fruitful discussions.
The work of M.N.~is supported in part by Grant-in-Aid for Scientific
Research (No.~20740141) from the Ministry
of Education, Culture, Sports, Science and Technology-Japan.

%%%%%%%%%%%%%%%%%%%%%%%%%%%%%%%%%%%%%%%%%%%%%%%%%%%%%%%%%%%

\end{document}